# An Overview of Wireless Local Area Networks
## (WLAN)

**Ibrahim Al Shourbaji**

**Computer Networks Department**
**Jazan University**
Jazan 82822-6649, Saudi Arabia

## Abstract

Wireless Communication is an application of science and technology that has come to be vital for modern existence. From the early radio and telephone to current devices such as mobile phones and laptops, accessing the global network has become the most essential and indispensable part of our lifestyle. Wireless communication is an ever developing field, and the future holds many possibilities in this area. One expectation for the future in this field is that, the devices can be developed to support communication with higher data rates and more security. Research in this area suggests that a dominant means of supporting such communication capabilities will be through the use of Wireless LANs. As the deployment of Wireless LAN increases well around the globe, it is increasingly important for us to understand different technologies and to select the most appropriate one .

This paper provides a detailed study of the available wireless LAN technologies and the concerned issues ,will give a brief description of what wireless LANs are ,the need of Wireless LAN ,History of wireless LAN , advantages of Wireless Networks ,with summarizing the related work on WLAN in academic area , Wireless LAN technologies , some risks attacks against wireless technologies , suggesting some recommendations to protect wireless LAN network from attack , Finally we propose some research issues should be focused on in the future .

*Keywords*: *Wireless Networking, Security, 802.11 Standard, Network security,*

## I. INTRODUCTION

Computer technology has rapidly growth over the past decade, Much of this can be attributed to the internet as many computers now have a need to be networked together to establish an online connection. As the technology continues to move from wired to wireless, the wireless LAN (local area network) has become one of the most popular networking environments.

Companies and individuals have interconnected computers with local area networks (LANs).The LAN user has at their disposal much more information, data and applications than they could otherwise store by themselves. In the past all local area networks were wired together and in a fixed location. Wireless technology has helped to simplify networking by enabling multiple computer users simultaneously share resources in a home or business without additional or intrusive wiring.

The increased demands for mobility and flexibility in our daily life are demands that lead the development

## 2. What is a WLAN ?

To know WLAN we need first to know the definition of LAN, which is simply a way of connecting computers together within a single organization, and usually in a single site (Franklin, 2010).

According to Cisco report in 2000 wireless local-area network (WLAN) does exactly what the name implies: it provides all the features and benefits of traditional LAN technologies such as Ethernet and Token Ring without the limitations of wires or cables. Obviously, from the definition the WLAN is the same as LAN but without wires.

(Clark et al, 1978) defined WLAN as a data communication network, typically a packet communication network, limited in geographic scope.' A local area network generally provides high-bandwidth communication over inexpensive transmission media.

While (Flickenger, 2005) see it as a group of wireless access points and associated infrastructure within a limited geographic area, such as an office building or building campus, that is capable of radio communications. Wireless LANs are usually implemented as extensions of existing wired LANs to provide enhanced user mobility.

Wireless Local Area Network (WLAN) links two or more devices using a wireless communication method. It usually provides a connection through an Access Point (AP) to the wider internet (Putman, 2005).

This gives users the ability to move around within a local coverage area while still be connected to the network. Just as the mobile phone frees people to make a phone call from anywhere in their home, a WLAN permits people to use their computers anywhere in the network area.

In WLAN Connectivity no longer implies attachment. Local areas are measured not in feet or meters, but miles or kilometers. An infrastructure need not be buried in the ground or hidden behind the walls, so we can move and change it at the speed of the organization.

# 3. Why would anyone want a wireless LAN?

There are many reasons: (perm, 2000)

**1- An increasing number of LAN users are becoming mobile.** These movable users require that they are connected to the network regardless of where they are because they want simultaneous access to the network. This makes the use of cables, or wired LANs, impractical if not impossible.

**2- Wireless LANs are very easy to install.** There is no requirement for wiring every workstation and every room. This ease of installation makes wireless LANs inherently flexible. If a workstation must be moved, it can be done easily and without additional wiring, cable drops or reconfiguration of the network.

**3- Another advantage is its portability**. If a company moves to a new location, the wireless system is much easier to move than ripping up all of the cables that a wired system would have snaked throughout the building.
Most of these advantages also translate into monetary savings.

# 4. History of WLAN

(Negus & Petrick, 2009)
The wireless local area network (WLAN) is today everywhere device often taken for granted as a default interface for networked devices by users and manufacturers alike. But not very long ago, it was most definitely not so.

In the early 1990's WLANs found almost no success in selling to enterprise or campus environments as wired LAN replacements or enablers of mobility. The WLAN products of that day were far too slow, too expensive, too bulky, and too power hungry. Furthermore, mobile network connectivity was simply not yet a killer application. The "survivor" companies of that age were the ones who focused on adapting WLAN technology to specialty niches such as retailing, hospitality, and logistics.

Organizations that went after the "big" market of enterprise networking, and there were many that did, either went bankrupt or became largely scaled back divisions of large companies.

By the middle of the 1990's the WLAN industry had mainly consolidated into 4 players, But in the late 1990's the first significant market opportunity for WLANs emerged and it was quite unlike what the WLAN industry to date had largely envisioned.

The opportunity was the sharing of a broadband Internet connection within the home amongst multiple networked devices such as PCs initially, but inevitably also voice over Internet protocol (VoIP) phones, gaming consoles, media streamers and home automation appliances. Consumers, not enterprise IT managers, became the ones to choose what WLAN technology and products would achieve the de facto standard for the decade to follow.

## Advantages of Wireless Networks

Wireless LANs designed to operate in license-free bands making their operation and maintenance costs less than contemporary cellular and PCS networks. The use of license-free spectrum, however, increases the risk of network security and in-band interference. The key advantages of wireless networks as opposed to wired networks are mobility, flexibility, ease of installation and maintenance, and reduced cost. (Aziz, 2003)

According to (Symantec , 2002) wireless LANs are less expensive and less intrusive to implement and maintain, as user needs change.
Simple implementation and maintenance, extended reach,, increased worker mobility and reduce total cost of ownership and operation.

## Emerging Developments

Fundamental step forward in information theory, which first emerged during the time of the early development of wireless LANs, have now reached a level of maturity and acceptance that is allowing them to drive the quest for higher spectral efficiencies and data rates.

Another important development in wireless LAN technology is the emergence of mesh networking. Mesh networks have the potential to dramatically increase the area served by a wireless network. Mesh networks even have the potential, with sufficiently intelligent routing algorithms to boost overall spectral efficiencies attained by selecting multiple hops over high capacity links rather than single hops over low capacity links (Holt, 2005).

## 5-Wireless LAN Technologies

When making a decision about the best protocol or standard to use. We need to consider its features and our needs. Weight the features and compare the advantages and disadvantages of each one to make the final decision.

There are several wireless LAN solutions available today, with varying levels of standardization and interoperability. Many solutions that currently lead the industry, IrDa, Bluetooth, HomeRF and IEEE 802.11.  These technologies enjoy wider industry support and targeted to solve Enterprise, Home and public wireless LAN needs.

- **Infrared** (IrDa)

 The appearance of portable information terminals in work and living environments is increase the introduction of wireless digital links and local area networks(LAN's).

Wireless LANs can use either radio frequencies or infrared light to transmit signals. While it is considerably cheaper to install infrared networks, as many devices already have infrared (IrDA) ports (Franklin, 2010).

Portable terminals should have access to all of the services that are available on high-speed wired networks. Unlike their wired counterparts, portable devices are subject to severe limitations on power consumption, size and weight. The desire for inexpensive, high-speed links satisfying these requirements has motivated recent interest in infrared wireless communication (Gfeller & Bapst, 1979).

Wireless infrared communications refers to the use of free-space propagation of light waves in the near infrared band as a transmission medium for communication (Carruthers, 2002).

The Infrared Data Association (IrDA) is another trade association, which defined standards for infrared communication for many years. It has some advantages; notably that it is cheap and there are many devices which already include infrared including most laptops and PDAs as well as some printers. Before the advent of radio frequency LANs people were building infrared LANs, with some success. (irda.org, 2011)

The wavelength band between about 780 and 950 nm is presently the best choice for most applications of infrared wireless links, due to the availability of low-cost LED's and laser diodes (LD's), and because it coincides with the peak responsively of inexpensive, low-capacitance silicon photodiodes (Rancourt,, 1993).

It provide a useful complement to radio-based systems, particularly for systems requiring low cost, light weight, moderate data rates, and only requiring short ranges (Carruthers, 2002).

However, this radiation cause problem relates to eye safety; it can pass through the human cornea and focused by the lens onto the retina, where it can potentially induce thermal damage (Kahn & Barry, 1997).

To achieve eye safety with an LD user can employ a thin plate of translucent plastic. such diffusers can achieve efficiencies of about 70%, offering the designer little freedom to tailor the source radiation pattern. Computer generated holograms (Smyth et al, 1995).

The primary goals in extending IrDA-Data's connection model were: (Williams, 1999)

- To enable devices to view each other to establish communication relationships uninhibited by the connection state of nearby devices.

- To enable an AIR device to establish communications with at most one IrDA 1.x device.

- For AIR devices to respect established connections with which they could interfere. This is a co-existence requirement intended to ensure that AIR devices do not disrupt active connections

## • Bluetooth

Bluetooth is an industry specification for short-range connectivity for portable personal devices with its functional specification released out in 1999 by Bluetooth Special Interest Group.
Bluetooth communicates on a frequency of 2.45 gigahertz, which has been set aside by international agreement for the use of industrial, scientific and medical devices (ISM) (Chandramouli, 2005). It is a worldwide license free band that any system can use (Goldsmith, 2004).

Using this band allows the Bluetooth protocol to become a standard around the world for interfacing devices together wirelessly.
Communications protocol developed to allow the devices using Bluetooth to transfer data reliably over their wireless network.

Bluetooth has a range of less than 10 meters. The range is increased when a scatternet is used because each unit only has to be within 10 meters of one other unit. The range can also be increased if the data is transmitted in a high power mode which offers transmissions up to 100 meters. Bluetooth also offers a cipher algorithm for security. This is most useful in the high power mode because when data is being transmitted further there is a greater possibility of an unwanted device receiving the network's data (Goldsmith, 2004).

## • HomeRF

In early 1997, several companies formed the Home RF working group to begin the development of a standard designed specifically for wireless voice and data networking in the home. (Goldsmith, 2004). HomeRF is an open industry specification developed by Home Radio Frequency Working Group (Wireless Networking Choices for the Broadband Internet Home., 2001) that defines how electronic devices such as PCs, cordless phones and other peripherals share and communicate voice, data and streaming media in and around the home.

The development of this working group was motivated by the widespread use of the internet and the development of affordable PCs that can be used in most homes. This protocol allows PCs in the home to have greater mobility, providing a connection to the Internet, printers, and other devices anywhere in the home. With all this potential, many members of industry worked to develop the Shared Wireless

Access Protocol-Cordless Access (SWAP-CA) specification (Goldsmith, 2004).

Unlike Wi-Fi, HomeRF already has quality-of-service support for streaming media and is the only wireless LAN to integrate voice. HomeRF may become the worldwide standard for cordless phones. In the year 2001, the Working group unveiled HomeRF 2.0 that supports 10 Mbps (HomeRF 2.0) or more. (Chandramouli, 2005)

A network topology of the Home RF protocol consists of four types of nodes: Control Point, Voice Terminals, Data Nodes, and Voice and Data Nodes. The control point is the gateway to the public switched telephone network (PSTN) and the Internet. It is also responsible for power management of the network. A voice terminal communicates with the control point via voice only. A data node communicates with the control point and other data nodes. Finally, a voice and data node is a combination of the previous two nodes (Lansford, 2000).

## • IEEE 802.11

The vendors joined together in 1991, first proposing, and then building, a standard based on contributed technologies. In June 1997, the IEEE released the 802.11 standard for wireless local-area networking (Cisco Wireless Lan standard report, 2000).

This initial standard specifies a 2.4 GHz operating frequency with data rates of 1 and 2 Mbps. With this standard, one could choose to use either frequency hopping or direct sequence. Because of relatively low data rates as, products based on the initial standard did not flourish as many had hoped (Chandramouli, 2005).

In late 1999, the IEEE published two supplements to the initial 802.11 standard: 802.11a and 802.11b (Wi-Fi). The 802.11a (Highly Scalable Wireless LAN Standard , 2002) standard (High Speed Physical Layer in the 5 GHz Band) specifies operation in the 5 GHz band with data rates up to 54 Mb/s (O'Hara, B. and Petrick, 1999).

The 802.11 WLAN standard allows for transmission over different media. Compliant media include infrared light and two types of radio transmission within the unlicensed 2.4-GHz frequency band: frequency hopping spread spectrum (FHSS) and direct sequence spread spectrum (DSSS).
Spread spectrum is a modulation technique developed in the 1940s that spreads a transmission signal over a broad band of radio frequencies.

Several studies talk about protocols and its characteristics, all the protocols developed for their own specific needs and they are capable of filling these needs well.

We will mention some of them briefly in a table according to (Goldsmith, 2004) study.

| Characteristic | Bluetooth | HomeRF |
|---|---|---|
| Operational Spectrum | 2.402 - 2.480 GHz | 2.404 - 2.478 GHz |
| Bandwidth | 78 MHz | 74 MHz |
| Modulation Type | FHSS (1600 Hops/sec), GFSK | FHSS (50 Hops/sec), 2-FSK, 4-FSK |
| Channel Access | Master-Slave Polling | CSMA/CA and TDMA |
| Data Rates | .721 Mbps Peak | .8, 1.8 Mbps |
| Data Traffic | PPP | TCP/IP |
| Range | Regular – 10 m High Power – 100 m | 50 m |
| Error Robustness | 1/3 rate FEC, 2/3 rate FEC, ARQ Type 1 | CRC/ARQ Type I |
| Security | YES | YES |
| Communications Topology | Peer-to-Peer, Master-to-Slave | Peer-to-Peer, MS-to-BS |
| Vender Stability | Very Good | N/A |
| Device Scalability | Currently Very Low | Good |
| Data Scalability | Low | OK |
| Transmit Power | NA | 100 mW |
| Energy Conservation | Yes | Directory Based |
| Capital Cost | Adapter: ~$30 Chipset: Under $4 in Bulk | N/A |
| Operational Cost | None | N/A |

Wireless security has become just as important as the technology itself. This issue known in the media with much press on how easy it is to gain unauthorized access to a wireless network. It seems as if this attention has fallen on deaf ears as these networks are still incredibly in danger. The absence of a physical connection between nodes makes the wireless links vulnerable to spy and information theft.

Insecure wireless user stations such as laptops create an even greater risk to the security of the enterprise network than rogue access points. The default configuration of these devices offer little security and can be easily misconfigured. Intruders can use any insecure wireless station as a launch pad to break in the network.

The basis for all WLAN security should start by understanding the environment in which your WLAN operates and its benifits.

We think about mobility and productivity as benefits of wireless, but that benefits put your information at risk.

We should pay attention on security alerts and set up a secure WLANs by implementing some practical actions.

(Khatod, 2004) implement five steps to protect the information assets, identify vulnerabilities and protect the network from wireless-specific attacks.

1. Discovery and improvement of Unauthorized WLANs and Vulnerabilities.
   it represent one of the biggest threats to enterprise network security by creating an open entry point to the enterprise network that bypasses all existing security measures including access points, soft access points (laptops acting as access points), user stations, wireless bar code scanners and printers.

   According to wireless security experts, discovery of unauthorized access points, stations and vulnerabilities is best accomplished with full monitoring of the WLAN.

2. Lock Down All Access Points and Devices
   The next step of WLAN security involves perimeter control for the WLAN. Each wireless equipped laptop should be secured by deploying a personal agent that can alert the enterprise and user of all security vulnerabilities and enforce conformance to enterprise policies. Organizations should deploy enterprise-class access points that offer advanced security and management capabilities.

3. Encryption and Authentication
   Encryption and authentication provide the core of security for WLANs. However ,fail-proo encryption and authentication standards have yet to be implemented.

4. Set and Enforce WLAN Policies

WLANs needs a policy for usage and security. While policies will vary based on individual security and management requirements of each WLAN, a thorough policy and enforcement of the policy can protect an enterprise from unnecessary security breaches and performance degradation.

5. Intrusion Detection and Protection

Security mangers rely on intrusion detection and protection to ensure that all components of WLANs are secure and protected from wireless threats and attacks.

To avoid the risks we should know it first, understanding how they work and using this information to avoid them as a solution for WLANs security.

A report from Internet Security Systems incorporation discuss some risks attacks against wireless technologies, they fall into seven basic categories:

1. Insertion attacks
2. Interception and unauthorized monitoring of wireless traffic
3. Jamming
4. Client-to-Client attacks
5. Brute force attacks against access point passwords
6. Encryption attacks
7. Misconfigurations

### 1- Insertion Attacks

Insertion attacks are based on deploying unauthorized devices or creating new wireless networks without going through security process and review (Bidgoli, 2006).

### 2- Interception and Monitoring of Wireless Traffic

As in wired networks, it is possible to intercept and monitor network traffic across a wireless LAN.

The attacker needs to be within range of an access point (approximately 300 feet for 802.11b) for this attack to work, The advantage for a wireless interception is that a wired attack requires the placement of a monitoring agent on a compromised system. All a wireless intruder needs is access to the network data stream.

### 3- Jamming

jamming can be a massive problem for WLANs. It is one of many exploits used compromise the wireless environment. It works by denying service to authorized users as legitimate traffic is jammed by the overwhelming frequencies of illegitimate traffic.

### 4- Client-to-Client Attacks

Two wireless clients can talk directly to each other, bypassing the access point. Users therefore need to defend clients not just against an external threat but also against each other.

### 5- Brute Force Attacks Against Access Point Passwords

Most access points use a single key or password that is shared with all connecting wireless clients. Brute force dictionary attacks attempt to compromise this key by methodically testing every possible password. The intruder gains access to the access point once the password is guessed.

### 6- Attacks against Encryption

802.11b standard uses an encryption system called WEP (Wired Equivalent Privacy). WEP has known weaknesses (see http://www.isaac.cs.berkeley.edu/isaac/wep-faq.html for more information), and these issues are not slated to be addressed before 2002. Not many tools are readily available for exploiting this issue, but sophisticated attackers can certainly build their own.

### 7- Misconfiguration

Many access points ship in an unsecured configuration in order to emphasize ease of use and rapid deployment. Unless administrators understand wireless security risks and properly configure each unit prior to deployment, these access points will remain at a high risk for attack or misuse.

Another report about Securing Wireless Local Area Networks suggests recommendations to protect wireless LAN network from attack, the following are some of them:

1. Educate employees about WLAN risks, and how to recognize an intrusion or suspicious behavior.
2. restrict unauthorized attachment of wireless access points (rogue access points).
3. Employ a third party managed security services company to constantly monitor the network security infrastructure for signs of an attack or unauthorized use.
4. Deploy strong for all of IT resources.
5. Ask users to connect only to known access points; masquerading access points are more likely in unregulated public spaces.
6. Deploy personal firewalls, anti-virus software and spyware blockers on all corporate PCs, particularly laptops and computers using the Windows operating system.
7. Actively and regularly scan for rogue access points and vulnerabilities on the corporate network, using available WLAN management tools.
8. Change default management passwords and, where possible, administrator account names, on WLAN access points.

9. Use strong security for other data resources such as laptop or desktop data files and e-mail messages and attachments.

10. Avoid placing access points against exterior walls or windows.

11. Reduce the broadcast strength of WLAN access points, when possible, to keep it within the necessary area of coverage only.

12. Using of an Intrusion Detection System. This will provide your wireless network with early detection of common threats

**Future works**

Future work should focus on the following issues:

- Lack of method to detect a passive sniffer: An attacker usually first collects data traffic before launching an intrusion. This type of passive sniffing is quite dangerous, but there is nothing to do in this direction except to use the proper protection through encryption.

- To think about how to reduce and eliminate the risks attacks that can be happened on WLAN networks such as Man-in-the Middle attacks , Denial of Service (DoS) attacks and Identity theft (MAC spoofing)

- Authentication is the key: The most significant vulnerability of wireless LANs is the fact that, at the physical level, by definition they enable access to anyone, authorized or not, within a WLAN access point's radius of useful signal strength.

# Conclusion

The future of wireless local-area networking is now, and it is the solution for communication problems in organizations or any place that need a wide spread of internet connection , interoperability became reality with the introduction of the standards and protocols and prices have dramatically decreased. These improvements are just a beginning.

Organizations who use WLANs networks can eliminate many of wireless LAN's security risks with careful education, planning, implementation, and management.

WLAN brings out not only advantages, but also some Specific security problems, although development of wireless standards and security protocols may enhance the WLAN security.

We know that hackers will never go away, so we bear the burden to provide the best 'locks' we can to protect our WLANs.  Finally, whatever the outcome, wireless LANs will survive and are here to stay even if the technology has a new look and, or feel in coming years.